\documentclass[aps,pra,floatfix,superscriptaddress,twocolumn,showpacs,showkeys,10pt]{revtex4-1}
\usepackage{amssymb, amsmath,color,mciteplus,graphicx,subfigure}
\usepackage{calc,epsfig,epstopdf,color,mciteplus,bm,mathrsfs}
\usepackage{times}
\usepackage{float}
\usepackage{lipsum}

\begin{document}

\title{Dynamical-Corrected Nonadiabatic Geometric Quantum Computation}

\author{Cheng-Yun Ding}
\affiliation{School of Mathematics and Physics, Anqing Normal University, Anqing 246133, China}
\affiliation{Key Laboratory of Atomic and Subatomic Structure and Quantum Control (Ministry of Education),\\ and School of Physics, South China Normal University, Guangzhou 510006, China}

\author{Li Chen}
\affiliation{School of Mathematics and Physics, Anqing Normal University, Anqing 246133, China}

\author{Li-Hua Zhang}\email{zhanglh@aqnu.edu.cn}
\affiliation{School of Electronic Engineering and Intelligent Manufacturing, Anqing Normal University, Anqing 246133, China}
\affiliation{School of Mathematics and Physics, Anqing Normal University, Anqing 246133, China}

\author{Zheng-Yuan Xue}\email{zyxue83@163.com}
\affiliation{Key Laboratory of Atomic and Subatomic Structure and Quantum Control (Ministry of Education),\\ and School of Physics, South China Normal University, Guangzhou 510006, China}

\affiliation{Guangdong Provincial Key Laboratory of Quantum Engineering and Quantum Materials,\\    Guangdong-Hong Kong Joint Laboratory of Quantum Matter,  and Frontier Research Institute for Physics,\\ South China Normal University, Guangzhou 510006, China}

\date{\today}

\begin{abstract}
Recently, nonadiabatic geometric quantum computation has been received great attentions, due to its fast operation and intrinsic error resilience. However, compared with the corresponding dynamical gates, the robustness of implemented nonadiabatic geometric gates based on the conventional single-loop geometric scheme still has the same order of magnitude due to the requirement of strict multi-segment geometric controls, and the inherent geometric fault-tolerance characteristic is not fully explored. Here, we present an effective geometric scheme combined with a general dynamical-corrected technique, with which the super-robust nonadiabatic geometric quantum gates can be constructed over the conventional single-loop geometric and two-loop composite-pulse geometric strategies, in terms of resisting the systematic error, i.e., $\sigma_x$ error. In addition, combined with the decoherence-free subspace (DFS) coding, the resulting geometric gates can also effectively suppress the $\sigma_z$ error caused by the collective dephasing. Notably, our protocol is a general one with simple experimental setups, which can be potentially implemented in different quantum systems, such as Rydberg atoms, trapped ions and superconducting qubits. These results indicate that our scheme represents a promising way to explore large-scale fault-tolerant quantum computation.
\end{abstract}

\maketitle

\section{Introduction}
In the last several decades, quantum computation has been gradually received great attentions as it has stronger computing power than classical computation and can solve certain problems that cannot be solved by classical one. For example, Shor's and Grover's algorithms are the two most typical quantum algorithms in the early stage, for verifying the quantum computing capability \cite{shor1999polynomial,PhysRevLett.79.325}. Meanwhile, quantum computation has been theoretically proposed and experimentally verified in many quantum systems, such as nuclear magnetic resonance \cite{cory1997ensemble,gershenfeld1997bulk}, ion trap \cite{PhysRevLett.74.4091,PhysRevX.8.021012}, neutral atoms \cite{PhysRevA.72.032333,PhysRevLett.104.010503}, photons \cite{knill2001scheme,PhysRevLett.120.205062,zhong2020quatunm,su2022efficient} and superconducting circuit systems \cite{PhysRevLett.79.2371,makhlin1999josephson,nakamura1999coherent,
	nakamura1999coherent,PhysRevA.76.042319,devoret2013superconducting,arute2019quantum,PhysRevLett.127.180501}. However, in most of these experiments, the performances of the implemented quantum gates are still greatly affected by the local artificial-manipulation and random errors, resulting in the relatively low gate-fidelity, and thus it is still difficult to realize large-scale fault-tolerant quantum computation.

Geometric quantum computation \cite{ekert2000geometric,zanardi1999holonomic} is a kind of special quantum computation strategy that uses geometric phases \cite{berry1984quantal,PhysRevLett.52.2111,PhysRevLett.58.1593,anandan1988non} to construct universal quantum gates, and is considered as one of the promising alternatives to realize universal quantum computation. Since the geometric phase has intrinsic geometric property, that is, it does not change with the evolution details of the evolution state, and only depends on the global property of the evolution path, the geometric quantum gates based on the geometric phases can naturally resist the local systematic error, also called $\sigma_x$ error \cite{PhysRevLett.91.090404,PhysRevA.70.042316,PhysRevLett.94.100502,PhysRevLett.102.030404,
	PhysRevA.84.042335,solinas2012stability,PhysRevA.86.062322,PhysRevA.87.060303}. The geometric phase was proposed by Berry in $1984$ \cite{berry1984quantal}. However, since this phase is induced by adiabatic evolution of the system energy eigenstates, it requires that the system evolution must be slow enough, and thus being seriously affected by decoherence. Subsequently, the adiabatic Berry phase is extended to the nonadiabatic Aharonov-Anandan (A-A) phase \cite{PhysRevLett.58.1593}. It is not limited by the adiabatic theorem \cite{PhysRevLett.95.110407,PhysRevLett.104.120401}, so its fast evolution time makes the constructed geometric gates quickly realized in experiments \cite{leibfried2003experimental,song2017continuous,PhysRevLett.124.230503,zhao2021experimental}. Furthermore, Berry phase is also quickly extended to adiabatic and nonadiabatic non-Abelian geometric phases \cite{PhysRevLett.52.2111,anandan1988non}. The constructed quantum gates by the latter two are also generally called the holonomic quantum gates \cite{zanardi1999holonomic,sjoqvist2012non,PhysRevLett.109.170501}.

Recently, finding a short and robust evolution path is one of the important concerns to realize faster and more robust geometric quantum gates \cite{PhysRevA.96.052316,PhysRevApplied.10.054051,PhysRevA.101.052302,PhysRevA.103.032609,PhysRevApplied.14.064009,PhysRevResearch.2.023295,ji2021noncyclic,li2021high,ding2021nonadiabatic,ding2021path,PhysRevApplied.14.034038,PhysRevApplied.18.014062,PhysRevApplied.17.034015}. Simultaneously, it has been proposed to combine some positive pulse-optimized techniques, such as composite pulse \cite{PhysRevA.80.024302,PhysRevA.90.012341,PhysRevA.92.022333,PhysRevA.95.032311,liu2021realization} and dynamical decoupling technologies \cite{PhysRevLett.82.2417,PhysRevA.90.022323,PhysRevA.102.032627}, to suppress the errors during the implementation of the quantum gate. The geometric schemes based on the single-loop orange-slice-shaped evolution path have been proposed \cite{PhysRevA.96.052316,PhysRevApplied.10.054051,PhysRevA.101.052302,PhysRevA.103.032609} and experimentally demonstrated \cite{PhysRevLett.124.230503,zhao2021experimental} to realize universal geometric quantum gates. However, it requires strict control and connection of multi-segment pulses, causing the target gate-robustness against systematic $\sigma_x$ error without great advantages over the corresponding dynamical gates \cite{PhysRevApplied.10.054051}. Thus, to further improve the robustness against this error, the single-loop geometric scheme combined with the composite-pulse technique has been proposed \cite{PhysRevApplied.10.054051} and verified by experiment \cite{PhysRevApplied.12.024024}. Moreover, the more the number of increased composite pulses is, that is, the number of cycles of the evolution path is continuously increased, the stronger the robustness of constructed geometric gates resisting $\sigma_x$ error is. However, the greater the numbers turn, the longer the gate-operation time is, and thus the more seriously affected by the decoherence effect. So, considering the competition of the systematic $\sigma_x$ and decoherence errors, the geometric scheme based on two-loop composite pulse is the best choice, under the finite coherent time condition on superconducting circuits \cite{PhysRevApplied.10.054051}.

Here, to more furtherly improve the gate robustness compared to that using two-loop composite-pulse geometric scheme against $\sigma_x$ error, we apply the dynamical correction technique \cite{PhysRevLett.82.2417,PhysRevLett.102.080501,rong2015experimental} to the case of the conventional single-loop geometric gates, and propose a general scheme of universal dynamical-corrected nonadiabatic geometric quantum computation with a simple implementation. The nonadiabatic geometric and holonomic gates have different implementation ways, which are based on the nonadiabatic Abelian and non-Abelian geometric phases, respectively, with different energy-level structures. In addition, notice that in the corresponding holonomic scheme \cite{PhysRevApplied.16.044005}, it is necessary to add a dynamical evolution of rotating $\pi$ angle around the axis in the $xoy$ plane in the middle of each of the two evolution segments for the two-segment orange-slice-shaped path, and the dynamical phases generated by the two dynamical evolution just cancel each other. In the case of our geometric scheme, the difference is that we need to insert a dynamical evolution to the middle of each segments of the three evolution segments surrounding orange-slice-shaped path. Among them, only the inserted second-segment one is rotated by $\pi$ angle around the axis in the $xoy$ plane, while the inserted first- and third-segment ones are, respectively, rotated by a certain angle around the axis in the non-$xoy$ plane, and the sum of the two rotated angles is just equal to $\pi$. Besides, the sum of the three accumulated dynamical phases also needs to be equal to $0$. In this way, the constructed nonadiabatic geometric gates have much stronger robustness than the conventional single-loop geometric ones for $\sigma_x$ error. Numerical simulations show that the robustness of constructed dynamical-corrected geometric gates are both stronger than those of the corresponding conventional single-loop geometric and two-loop composite-pulse geometric ones for resisting $\sigma_x$ error. Moreover, combined with decoherence-free subspace (DFS) coding \cite{PhysRevLett.81.2594,kwiat2000experimental}, our scheme can not only resist $\sigma_x$ error, but also resist the $\sigma_z$ error caused by the collective dephasing. All these show that our scheme is a promising one for large-scale fault-tolerant quantum computation.

This paper is organized as follows. In Sec. \ref{sec2}, we briefly describe the conventional single-loop and composite-pulse geometric quantum gate schemes. The construction for dynamical-corrected geometric gates and comparison of gate performance for three different geometric schemes are presented in Sec. \ref{sec3}. The physical implementation of dynamically corrected geometric scheme combined with the DFS coding is made in Sec. \ref{sec4}, which includes single- and two-qubit cases, and the results are summarized in Sec. \ref{sec5}.

\section{conventional single-loop geometric and composite-pulse geometric gates}\label{sec2}
Because our dynamical-corrected geometric scheme is built on the conventional single-loop geometric scheme \cite{PhysRevA.96.052316,PhysRevApplied.10.054051}, here we briefly describe the construction of that and highlight the limitation for gate robustness compared with the corresponding dynamical counterpart in existing local $\sigma_x$ error. And, the composite-pulse geometric scheme is also described here, which can enhance the gate robustness of the conventional single-loop geometric one against $\sigma_x$ error.

\begin{figure}
\centering
\includegraphics[width=0.98\linewidth]{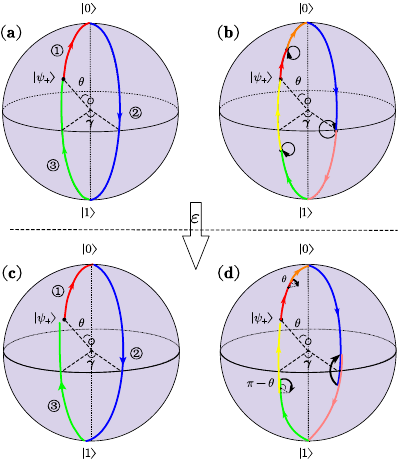}\\
\caption{Evolution trajectories for the conventional single-loop geometric scheme (a) without and (c) with $\sigma_x$ error and the dynamically corrected geometric one (b) without and (d) with $\sigma_x$ error on Bloch sphere.}\label{Figure1}
\end{figure}

For a driven two-level quantum system, its resonant Hamiltonian can be expressed as follows,  on the computational basis vectors $\{|0\rangle,|1\rangle\}$,
\begin{equation}\label{twolevelh}
\mathcal{H}(t)=\frac{\Omega(t)}{2}\left(
                            \begin{array}{cc}
                              0 & e^{-i\phi} \\
                              e^{i\phi} & 0 \\
                            \end{array}
                          \right),
\end{equation}
where $\Omega(t)$ and $\phi$ denote the driving coupling and initial phase of driving field, respectively. By dividing the evolution path of initial state $|\psi_+\rangle=\cos{(\theta/2)}|0\rangle+\sin{(\theta/2)}e^{i\phi}|1\rangle$ into three segments and setting the appropriate Hamiltonian parameters $\{\Omega(t),\phi\}$ in each evolution segment, a set of universal single-qubit geometric gates can be obtained.  Concretely, in the first segment, letting $\textstyle\int_0^{\tau_1}\Omega(t)dt=\theta$, which makes that the evolution state travels along the longitude line $\phi$ from starting point $(\theta,\phi)$ to the North Pole on the Bloch sphere; In the second segment, letting $\textstyle\int_{\tau_1}^{\tau_2}\Omega(t)dt=\pi$, then the evolution state follows another longitude line $\phi+\gamma$ to the South Pole; In the last segment, $\textstyle\int_{\tau_2}^{\tau_3}\Omega(t)dt=\pi-\theta$ must be satisfied, which ensures that the evolution state follows the longitude line $\phi$ from the South Pole returning to the initial point $(\theta,\phi)$ and completes the cycle evolution. The complete evolution trajectory on the Bloch sphere is shown in Fig. \ref{Figure1} (a), i.e., orange-slice-shaped path, in which we can find that only a geometric phase $\gamma$ is accumulated in the whole process. Similarly, the orthogonal evolution state $|\psi_-\rangle$ can also be accumulated a geometric phase $-\gamma$ after a cycle of evolution, and the final geometric evolution operator can be calculated as $U(\tau)=e^{i\gamma\textbf{n}\cdot\bm{\sigma}}$ with $\textbf{n}=(\sin{\theta}\cos{\phi},\sin{\theta}\sin{\phi},\cos{\theta})$ and $\bm{\sigma}=(\sigma_x,\sigma_y,\sigma_z)$, which represents the general single-qubit geometric gates.

When there is a $\sigma_x$ error, namely the form of pulse intensity $\Omega(t)$ changes as $\Omega(t) \rightarrow (1+\epsilon)\Omega(t)$ ($\epsilon$ is the error ratio for the pulse intensity), then the equation of pulse area in each segment cannot be strictly satisfied. This means that the evolution trajectory of the evolution state is destroyed, so that the final evolution state cannot return to the starting point and causing the gate infidelity, as shown in Fig. \ref{Figure1}(c). In addition, it has been shown that the gate infidelities against $\sigma_x$ error still have the same order of magnitude between the above single-loop geometric and corresponding dynamical ones \cite{PhysRevApplied.10.054051}. So, the robust geometric nature has not been fully proven. But fortunately, some studies have pointed out that the composite-pulse technique can improve the resistance of the geometric gate for $\sigma_x$ error. The specific process is as follows \cite{PhysRevA.80.024302,PhysRevA.95.032311}:

Defining $U_c(\gamma_c)$ is the geometric gate of iteration unit, where $\gamma_c=\gamma/N$. By continuously applying the iterative-element gate $N$ ($N\geq2$) times, the composite-pulse geometric gates can be obtained as
\begin{equation}
U_c(N\gamma_c)=[U_c(\gamma_c)]^N=e^{i\gamma\textbf{n}\cdot\bm{\sigma}}.
\end{equation}
For example, to get a geometric NOT gate ($\gamma=\pi/2$), we can apply the geometric iterative-element gate $U_c(\gamma_c)=\sqrt{X}$ twice in succession, in which we set $\gamma_c=\gamma/2=\pi/4$. Similarly, it can also be obtained by continuously applying three or more corresponding geometric iterative-element gates. However, as the number of iterations increases, the gate-operation time will also be prolonged multiply. When the decoherence effect is considered, the fidelity of the target geometric gates will drop sharply. So, we only consider the two-loop composite-pulse geometric one here for comparison in Sec. \ref{sec3}.
In the next section, we will describe the construction process of dynamically corrected nonadiabatic geometric gates by combining a general dynamical correction technique, whose robustness against $\sigma_x$ error is greatly improved, even better than that of two-loop composite-pulse geometric gates.
\begin{figure}
\centering
\includegraphics[width=0.98\linewidth]{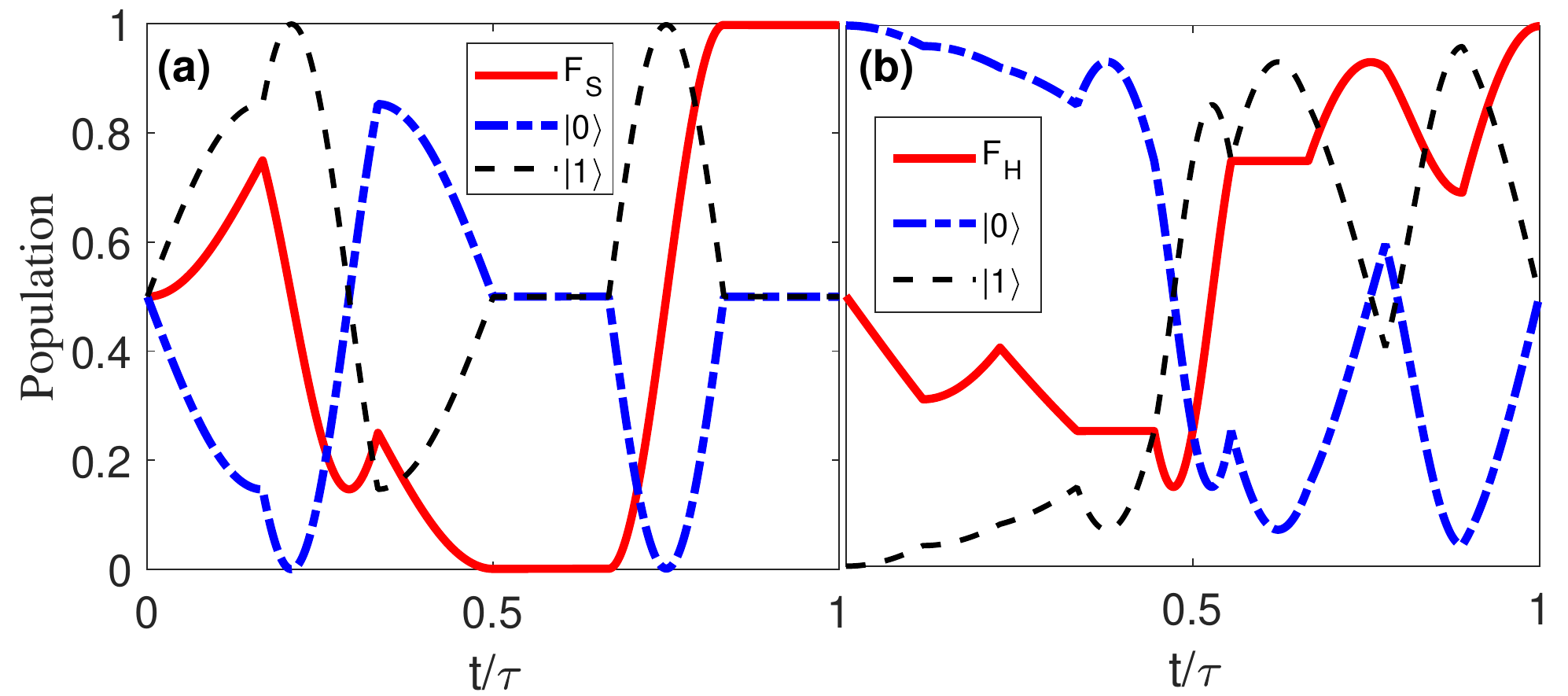}\\
\caption{Dynamics of state population and state fidelity for the dynamical-corrected geometric (a) S and (b) H gates, where the initial states are set on $(|0\rangle+|1\rangle)/\sqrt{2}$ and $|0\rangle$, respectively.}\label{Figure2}
\end{figure}

\section{Dynamical-corrected geometric gates}\label{sec3}
 To reduce the influence of $\sigma_x$ error on the fidelity of the conventional single-loop geometric gates, inspired by the dynamical correction technologies \cite{PhysRevLett.102.080501,rong2015experimental,PhysRevApplied.16.044005}, we insert a dynamical evolution in the middle of each of the three evolution segments for above-mentioned single-loop geometric scheme. Since the rotation axes of the inserted first- and third-segment dynamical evolutions are generally not in the $xoy$ plane, we need to consider a general two-level Hamiltonian, in which the detuning $\Delta(t)$ can be not $0$, that is
 \begin{equation}\label{DCGhamiltonian}
 \mathcal{H}_1(t)=\frac{1}{2}\left(
                            \begin{array}{cc}
                              \Delta(t) & \Omega(t)e^{-i\phi} \\
                              \Omega(t)e^{i\phi} & -\Delta(t) \\
                            \end{array}
                          \right).
\end{equation}

To obtain a set of universal dynamically corrected single-qubit geometric gates, we divide the whole evolution path into nine segments [in which the second, fifth and eighth segments are the inserted dynamical evolutions, and the corresponding evolution trajectories are plotted in Fig. \ref{Figure1} (b)], and then the Hamiltonian parameters $\{\Omega(t),\phi,\Delta(t)\}$ of each segment need to meet the following conditions:
\begin{equation}
\begin{split}
&\int_0^{\tau _1}{\Omega \left( t \right)}dt=\frac{\theta}{2},\ \phi -\frac{\pi}{2},\ \Delta =0;\
\\
&\int_{\tau _1}^{\tau _2}{\sqrt{\Omega ^2\left( t \right) +\Delta ^2\left( t \right)}}dt=\theta , \phi -2\pi ,\\ &\qquad\Delta(t) =\frac{\Omega(t)}{\tan(\theta/2)};
\\
&\int_{\tau _2}^{\tau _3}{\Omega \left( t \right)}dt=\frac{\theta}{2}, \phi -\frac{\pi}{2}, \Delta =0;
\\
&\int_{\tau _3}^{\tau _4}{\Omega \left( t \right)}dt=\frac{\pi}{2},\,\,\phi +\gamma +\frac{\pi}{2},\ \Delta =0;
\\
&\int_{\tau _4}^{\tau _5}{\Omega \left( t \right)}dt=\pi ,\,\,\phi +\gamma +\pi ,\ \Delta =0;
\\
&\int_{\tau _5}^{\tau _6}{\Omega \left( t \right)}dt=\frac{\pi}{2},\,\,\phi +\gamma +\frac{\pi}{2}, \Delta =0;
\\
&\int_{\tau _6}^{\tau _7}{\Omega \left( t \right)}dt=\frac{\pi -\theta}{2},\,\,\phi -\frac{\pi}{2},\,\,\Delta =0;\,\,
\\
&\int_{\tau _7}^{\tau _8}{\sqrt{\Omega ^2\left( t \right) +\Delta ^2\left( t \right)}}dt=\pi -\theta ,\,\,\phi -2\pi ,\\&\qquad\Delta(t) =\frac{\Omega(t)}{\tan[(\theta+\pi)/2]} ;
\\
&\int_{\tau _8}^{\tau _9}{\Omega \left( t \right)}dt=\frac{\pi -\theta}{2},\,\,\phi -\frac{\pi}{2},\,\,\Delta =0.\,\,
\end{split}
\end{equation}
At the end of evolution $\tau$, the corresponding evolution operator can be calculated as
\begin{equation}
\begin{split}
U(\tau)&=U(\tau_9,\tau_8)U(\tau_8,\tau_7)U(\tau_7,\tau_6)U(\tau_6,\tau_5)U(\tau_5,\tau_4)  \\
&\quad\times U(\tau_4,\tau_3)U(\tau_3,\tau_2)U(\tau_2,\tau_1)U(\tau_1,0) \\
&=e^{i\gamma\textbf{n}\cdot\bm{\sigma}}.
\end{split}
\end{equation}
In addition, note that the accumulated dynamical phases $\gamma_d^1$, $\gamma_d^2$ and $\gamma_d^3$ in the second, fifth and eighth segments can be calculated as $-\theta/2$, $\pi/2$ and $(\theta-\pi)/2$, respectively, whose sum is just zero, namely $\gamma_d^1+\gamma_d^2+\gamma_d^3=0$. Therefore, the target evolution operator $U(\tau)$ is still a pure geometric quantum gate without arbitrary dynamical phase. Next, we take a group of universal dynamical-corrected single-qubit geometric gates sequence (i.e., phase, $\pi/8$ and Hadamard gates, labelled as S, T and H) as the typical example to illustrate the gate performance of our scheme.
\begin{figure}
\centering
\includegraphics[width=0.98\linewidth]{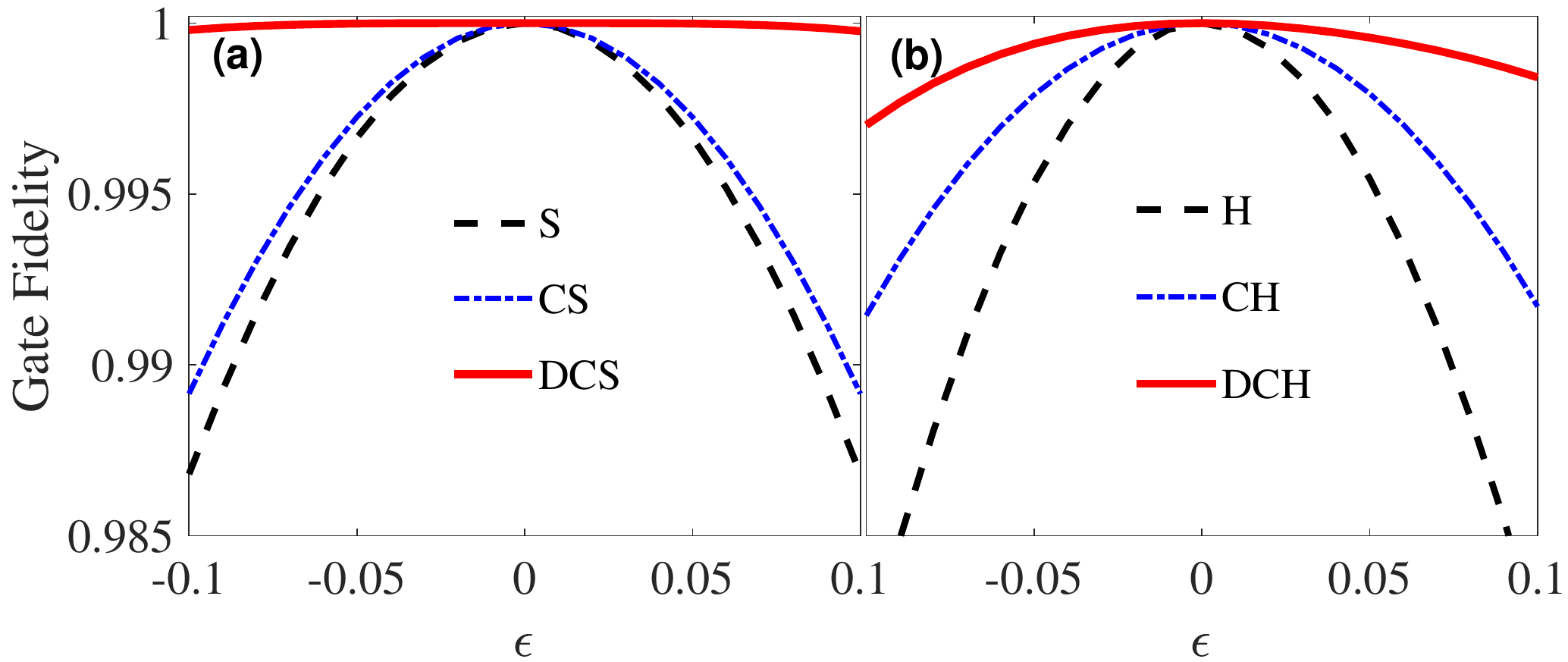}\\
\caption{Without decoherence, the gate robustness comparison against $\sigma_x$ error for the dynamical-corrected S (DCS), two-loop composite-pulse S (CS) and conventional single-loop S (S) geometric gates; (b) that of the corresponding H gates.}\label{Figure3}
\end{figure}

\begin{figure}
\centering
\includegraphics[width=0.98\linewidth]{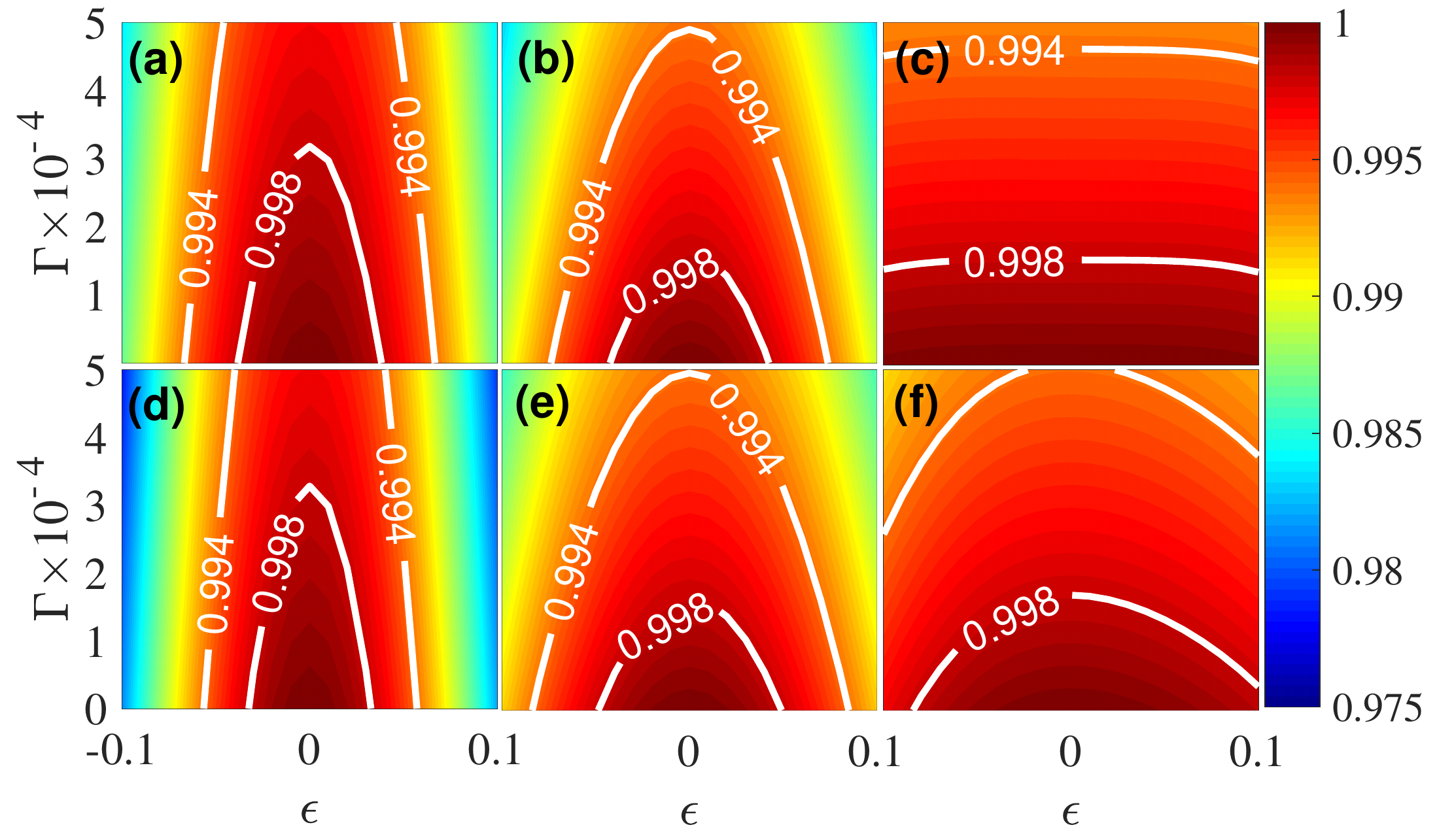}\\
\caption{For geometric S gates, the gate fidelity as a function of $\sigma_x$ error rate $\epsilon$ and decoherence rate $\Gamma$ based on (a) the conventional single-loop geometric, (b) two-loop composite-pulse geometric and (c) dynamical-corrected geometric schemes; For geometric H gates, the gate fidelity as a function of $\sigma_x$ error rate $\epsilon$ and decoherence rate $\Gamma$ based on (d) the conventional single-loop geometric, (e) two-loop composite-pulse geometric and (f) dynamical-corrected geometric schemes, where $\Gamma\in[0,5]\times10^{-4}$ and $\epsilon\in[-0.1,0.1]$.}\label{Figure4}
\end{figure}

To implement the corresponding S, T and H gates, we set the parameters $\theta$, $\phi$ and $\gamma$ to meet:
\begin{eqnarray}
\begin{split}
\left(\theta,\phi\right)&=\left(0,0\right),   \quad\quad\quad\! \gamma=-\pi/4,  \\
\left(\theta,\phi\right)&=\left(0,0\right),   \quad\quad\quad\! \gamma=-\pi/8, \\
\left(\theta,\phi\right)&=\left(\pi/4,0\right),   \quad\;\;\; \gamma=-\pi/2.
\end{split}
\end{eqnarray}
It is worth noting that since the $\theta=0$ must be satisfied for the implemented dynamical-corrected geometric S and T gates, our scheme can degenerate to the case similar to that of the dynamical-corrected  nonadiabatic holonomic quantum gates \cite{PhysRevApplied.16.044005}. That is, the whole evolution path only needs to be divided into six segments, and the detunings of the inserted two-stage symmetrical dynamical evolutions on the equator are also $0$. We utilizing Lindblad quantum master equation \cite{lindblad1976generators} of
\begin{equation}
\dot{\rho_1}=i[\rho_1,\mathcal{H}_1(t)]+\frac{1}{2}[\Gamma_1\mathcal{L}(\sigma_1)+\Gamma_2\mathcal{L}(\sigma_2)]
\end{equation}
to numerically simulate the dynamical process of the corresponding geometric gates, where $\rho_1$ is the density operator of the considered quantum system, $\mathcal{L}(\mathcal{A})=2\mathcal{A}\rho_1\mathcal{A}^{\dagger} -\mathcal{A}^{\dagger}\mathcal{A}\rho_1-\rho_1\mathcal{A}^{\dagger}\mathcal{A}$ is the Lindblad operator acting on operator $\mathcal{A}$, and $\sigma _1=\left| 0 \right> \left< 1 \right|$, $ \sigma _2=\left| 1 \right> \left< 1 \right|-\left| 0 \right> \left< 0 \right|$ and $\Gamma_1$, $\Gamma_2$ are the corresponding decay and dephasing rates, respectively. For convenience, in our numerical simulations, we set $\Omega(t)=\Omega_m=1$ and $\Gamma_1=\Gamma_2=\Omega_m/10^4$. As shown in Fig. \ref{Figure2}, we draw the dynamical process of the state population and state fidelity for the dynamically corrected geometric S and H gates (the gate performance of T gate is similar to that of  S gate, which will not be described hereafter), respectively. From the Fig. \ref{Figure2} we can find that the state fidelities for dynamically corrected geometric S and H gates can reach $99.86\%$ and $99.88\%$, respectively, in which the corresponding initial states $(|0\rangle+|1\rangle)/\sqrt{2}$ and $|0\rangle$ have been considered. As shown in Fig. \ref{Figure1}(d), when the three-segment dynamical-corrected pulses are inserted, the evolution states with $\sigma_x$ error can nearly return to the starting point eventually, and getting robust geometric quantum gates. To verify the gate robustness of our scheme, we consider the same form of $\sigma_x$ error $(1+\epsilon)\Omega(t)$. And, under this error, Eq. (\ref{DCGhamiltonian}) becomes
\begin{equation}
\mathcal{H}^E_1(t)=\frac{1}{2}\left(
                            \begin{array}{cc}
                              \Delta(t) & (1+\epsilon)\Omega(t)e^{-i\phi} \\
                              (1+\epsilon)\Omega(t)e^{i\phi} & -\Delta(t) \\
                            \end{array}
                          \right),
\end{equation}
where the error ratio $\epsilon$ is set in the interval $-0.1\leq\epsilon\leq0.1$. First, without considering decoherence, for geometric S and H gates, we compare the gate robustness for the conventional single-loop geometric, the two-loop composite-pulse geometric ones and our scheme, as shown in Fig. \ref{Figure3}. It can be seen that the gate robustness of our scheme is much stronger than those of the others, whose analytic solution is described in Appendix \ref{applxa}. Meanwhile, note that the total pulse areas of the geometric gates of our scheme are less than or equal to $4\pi$ (i.e., $\int_0^{\tau}\Omega(t)dt\leq4\pi$), and that of the two-loop composite-pulse geometric gates is $4\pi$, which are all larger than $2\pi$ for the conventional single-loop geometric scheme. Therefore, we need to consider the robustness comparison for different decoherence rates. Setting the decoherence rates as $\Gamma_1=\Gamma_2=\Gamma\in[0,5]\times10^{-4}$,  from the Fig. \ref{Figure4} we can see that the gate robustness for our scheme is still the best in the case of both considering $\sigma_x$ error and decoherence.

It is worth emphasizing that, for the parameter $\theta\neq0$, the robustness of our dynamical-corrected geometric gates are stronger than that of the recent similar optimization scheme resisting $\sigma_x$ error \cite{liang2022robust} (see the Appendix \ref{applxb} for details). Meanwhile, for the case of $\theta=0$, our optimization scheme can degenerate to the one in Ref. \cite{liang2022robust}. Thus, our dynamically corrected nonadiabatic geometric scheme is a more general one.

\section{Physical implementation}\label{sec4}
In the last section, we have demonstrated that the dynamical correction techniques can greatly improve the robustness of conventional single-loop geometric gates against $\sigma_x$ error. In this section, we will describe the physical implementation of our scheme combined with the DFS coding \cite{PhysRevLett.81.2594,kwiat2000experimental} and prove that such geometric gates can also effectively resist the $\sigma_z$ error caused by the collective dephasing.

\begin{figure}
\centering
\includegraphics[width=0.98\linewidth]{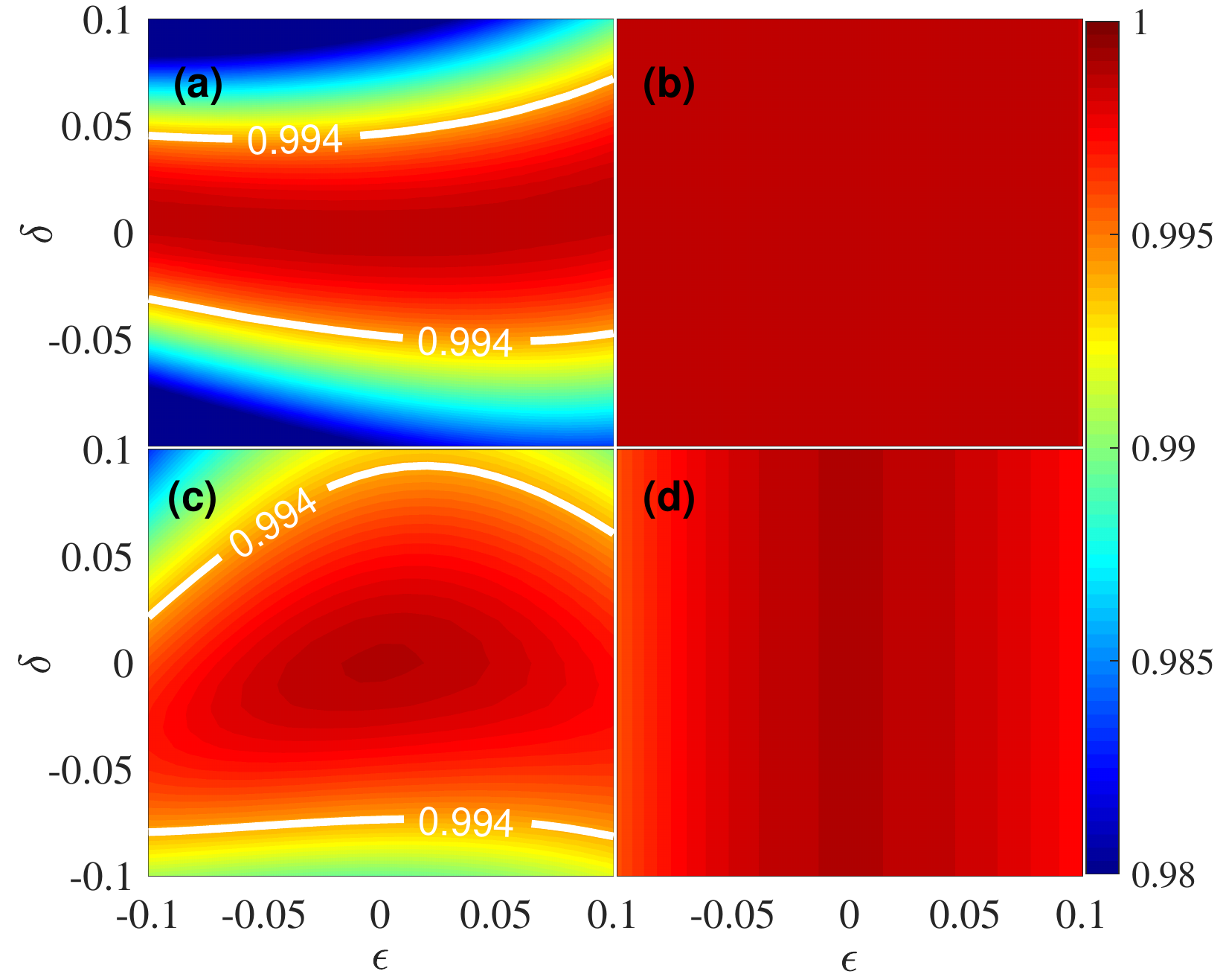}\\
\caption{For dynamical-corrected geometric S gate, the robustness comparison resists local $\sigma_x$ and $\sigma_z$ errors (a) without and (b) with DFS coding; For dynamical-corrected geometric H gate, the robustness comparison resists local $\sigma_x$ and $\sigma_z$ errors (c) without and (d) with DFS coding.}\label{Figure5}
\end{figure}

Considering two identical two-level physical qubits through direct-exchange interaction, their general Hamiltonian can be expressed as
\begin{equation}
\mathcal{H}_2\left( t \right) =\Delta \left( t \right) S'_z+\frac{J\left( t \right)}{2} \left[e^{-i\varphi}S_{1}^{+}S_{2}^{-}+\textrm{ H.c.}\right],
\end{equation}
where $\Delta(t)$ and $J(t)$ represent the detuning and coupling strength between physical qubits $1$ and $2$, respectively, H.c. denote the Hermite conjugate, $S'_z=\left(|1\rangle_{1}\langle1|-|1\rangle_{2}\langle1|\right)/2$, $S^{+}_{i}=|1\rangle_{i}\langle0|$ and $S^{-}_{i}=|0\rangle_{i}\langle1|$ with $i=1, 2$. Note that, there is a two-dimensional single-excitation DFS: $\mathcal{S}_1=\textrm{Span}\left\{ \left| 10 \right> _{12},\left| 01 \right> _{12} \right\} $, in which $|mn\rangle=|m\rangle_1\otimes|n\rangle_2$, and we can naturally encode the logical qubit as $\left| 0 \right> _L= \left| 10 \right> _{12}$ and $\left| 1 \right> _L=\left| 01 \right> _{12}$. Then, in the DFS $\mathcal{S}_1$, the above Hamiltonian can be rewritten into the following form:
\begin{equation}
\mathcal{H}_L\left( t \right) =\frac{1}{2}\left( \begin{matrix}
	{\Delta}\left( t \right)&		{J}\left( t \right) e^{-i\varphi}\\
	{J}\left( t \right) e^{i\varphi}&		-{\Delta}\left( t \right)\\
\end{matrix} \right).
\end{equation}
 It is easy to see that the form of the Hamiltonian $\mathcal{H}_L\left( t \right)$ is similar to that of the Hamiltonian $\mathcal{H}_1(t)$ for dynamical-corrected single-qubit geometric case without logical coding. Therefore, we can naturally realize the dynamical-corrected nonadiabatic geometric gates combined with DFS coding. In Fig. \ref{Figure5}, we test the robustness of dynamically corrected geometric S and H gates with and without DFS coding, respectively, to resist local $\sigma_x$ and $\sigma_z$ errors. Here, we define the Hamiltonian form influenced by these two kinds of errors as
 \begin{equation}
\begin{split}
\mathcal{H}_{L}^{E}\left( t \right) &=\frac{1}{2}{\Delta}\left( t \right) \sigma _{z}^{L}+\frac{{J}\left( t \right)(1+\epsilon)}{2}\big[ e^{-i\varphi}\left| 0 \right> _L\left< 1 \right|\\
&\quad+e^{i\varphi}\left| 1 \right> _L\left< 0 \right| \big] -\delta \Omega _m\left( \left| 0 \right> _L\left< 0 \right|+\left| 1 \right> _L\left< 1 \right| \right) .
\end{split}
\end{equation}
where $\delta\Omega_m$ denotes the collective dephasing errors, namely the Z errors are the same in each physical qubits. When this error is non-collective, the DFS coding can still eliminate the overlapping part of the Z errors.
In addition, in our numerical simulations, let all the coupling strengths be set to constant $1$, i.e., ${J\left(t\right)}=\Omega_m=1$, and the decoherence rates be set to $10^{-4}$. From Fig. \ref{Figure5}, we can see that the constructed geometric S and H gates can effectively resist $\sigma_x$ and $\sigma_z$ errors, simultaneously, after combining DFS coding.

To realize the universal quantum computation, a nontrivial two-qubit quantum gate is necessary. Similarly, we consider four identical two-level physical qubits, in which the first (third) and second (fourth) physical qubits are encoded as the first (second) logical qubit, and the effective Hamiltonian between them can be written as
\begin{equation}
\begin{split}
\mathcal{H}_3\left( t \right) &=\widetilde{\Delta} \left( t \right) S''_z+\frac{g\left( t \right)}{2}\Big[ e^{-i\widetilde{\varphi}}(|1\rangle_1\langle1|\otimes S_{2}^{+}S_{3}^{-}\otimes|0\rangle_4\langle0|   \\
&\quad+|0\rangle_1\langle0|\otimes S_{2}^{-}S_{3}^{+}\otimes|1\rangle_4\langle1| )+\textrm{H.c.}\Big],
\end{split}
\end{equation}
where $g(t)$ is the effective coupling strength between physical qubits $2$ and $3$. Considering a DFS coding for two-logical qubit, there is a $6$-dimensional DFS: $\mathcal{S}_2=\{|1010\rangle,|1001\rangle,|0110\rangle,|0101\rangle,|1100\rangle,|0011\rangle\}$, in which $|mnm'n'\rangle=|m\rangle_1\otimes|n\rangle_2\otimes|m'\rangle_3\otimes|n'\rangle_4$. Furthermore, the basis vectors of two-logical qubit is defined as $\{|00\rangle_L=|1010\rangle,|01\rangle_L=|1001\rangle,|10\rangle_L=|0110\rangle,|11\rangle_L=|0101\rangle\}$, and $|a_1\rangle_L=|1100\rangle$ and $|a_2\rangle_L=|0011\rangle$ are treated as two auxiliary states. Therefore, in the DFS for two-logical qubit $\mathcal{S}'_2=\{|00\rangle_L,|a_1\rangle_L,|01\rangle_L,|10\rangle_L,|11\rangle_L,|a_2\rangle_L\}$, the above Hamiltonian $\mathcal{H}_3(t)$ will become
\begin{equation}
\mathcal{H}^L_3(t)=\frac{1}{2}\left(
                     \begin{array}{cccccc}
                       -\widetilde{\Delta}(t) & g(t)e^{i\widetilde{\varphi}} & 0 & 0 & 0 & 0 \\
                       g(t)e^{-i\widetilde{\varphi}} & \widetilde{\Delta}(t) & 0 & 0 & 0 & 0 \\
                       0 & 0 & 0 & 0 & 0 & 0 \\
                       0 & 0 & 0 & 0 & 0 & 0 \\
                       0 & 0 & 0 & 0 & -\widetilde{\Delta}(t) & g(t)e^{i\widetilde{\varphi}} \\
                       0 & 0 & 0 & 0 & g(t)e^{-i\widetilde{\varphi}} & \widetilde{\Delta}(t)\\
                     \end{array}
                   \right).
\end{equation}
It is noted that the Hamiltonians in the two-logical-qubit subspaces $\{|00\rangle_L,|a_1\rangle_L\}$ or $\{|11\rangle_L,|a_2\rangle_L\}$ have a similar form as that of the single-logical-qubit case, so a same dynamically corrected geometric phase gate can be constructed in the corresponding logical subspaces with the detuning $\widetilde{\Delta}=0$, respectively. At the same time, by tracing the two auxiliary states, the dynamically corrected geometric gates for two-logical qubit can be obtained as follows:
\begin{equation}
U_2(\widetilde{\gamma})=\left(
                 \begin{array}{cccc}
                   e^{-i\widetilde{\gamma}} & 0 & 0 & 0 \\
                   0 & 1 & 0 & 0 \\
                   0 & 0 & 1 & 0 \\
                   0 & 0 & 0 & e^{-i\widetilde{\gamma}} \\
                 \end{array}
               \right).
\end{equation}

\begin{figure}
\centering
\includegraphics[width=1\linewidth]{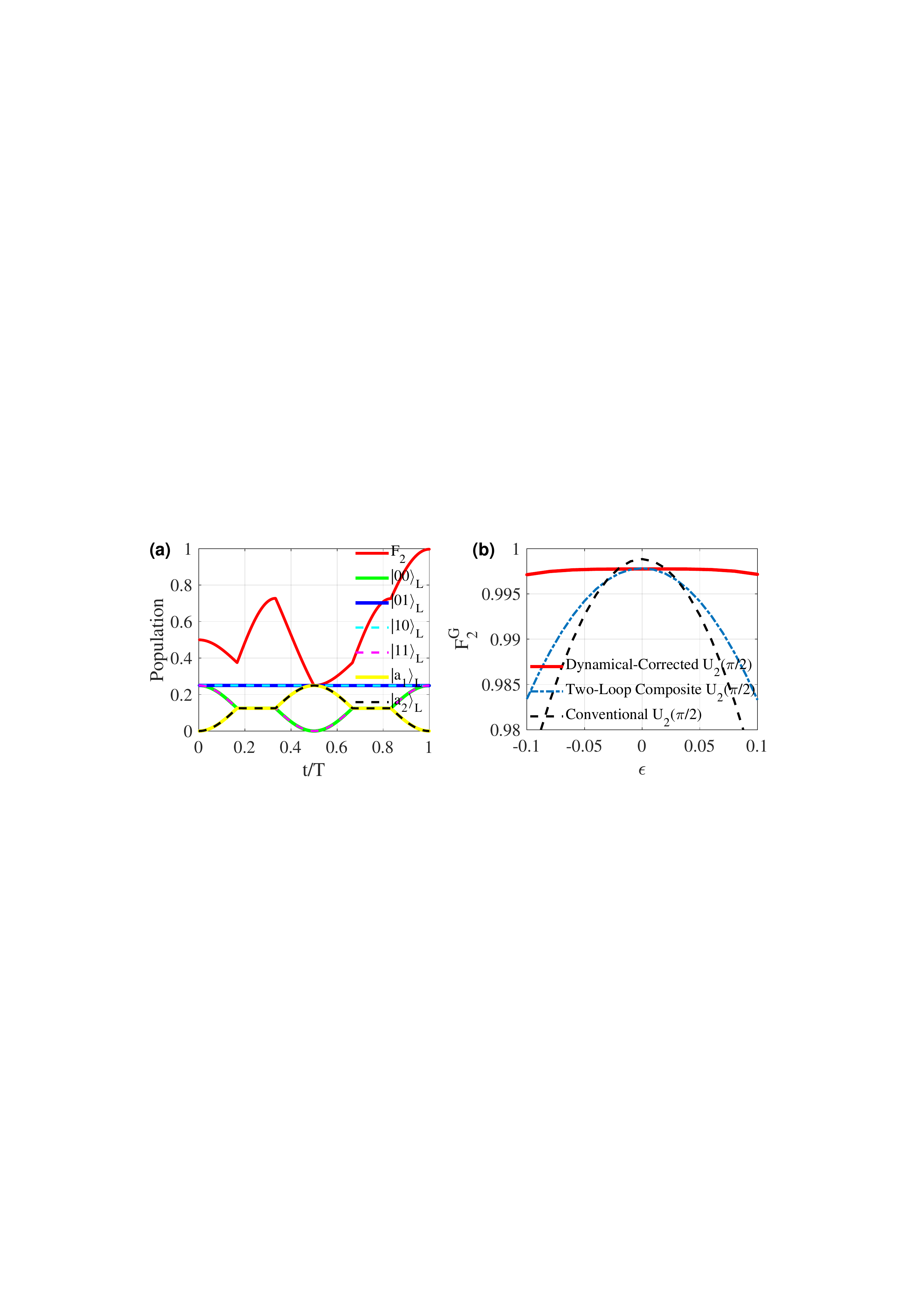}\\
\caption{The performance for dynamical-corrected two-logical-qubit geometric gate. (a) Dynamics of the state population and state fidelity for $U_2(\pi/2)$ with an initial state $|\psi_0\rangle=(|00\rangle_L+|01\rangle_L+|10\rangle_L+|11\rangle_L)/2$. (b) The robustness comparison of $U_2(\pi/2)$ against local $\sigma_x$ error for the conventional single-loop geometric, two-loop composite-pulse geometric and dynamical-corrected geometric schemes with DFS coding.}\label{Figure6}
\end{figure}

 With the help of several single-logical-qubit geometric gates, $U_2(\widetilde{\gamma})$ is actually equivalent to the nontrivial two-qubit geometric controlled-phase gates \cite{PhysRevA.105.012611}. Setting $\widetilde{\gamma}=\pi/2$, we simulate the state population and state fidelity $F_2$ for $U_2(\pi/2)$ based on the initial state $|\psi_0\rangle=\frac{1}{2}(|00\rangle_L+|01\rangle_L+|10\rangle_L+|11\rangle_L)$. As shown in Fig. \ref{Figure6} (a), the state fidelity $F_2$ can reach $99.67\%$ with the set decay and dephasing rates and coupling constants being the same as the ones of above single-logical-qubit geometric gates. Besides, to illustrate the robustness advantage for $U_2(\pi/2)$, we plot the robustness comparison against local $\sigma_x$ error among the conventional single-loop geometric, two-loop composite geometric and our dynamical-corrected geometric schemes in Fig. \ref{Figure6} (b).  The influenced two-logical-qubit Hamiltonian in the DFS $\mathcal{S}'_2$ can be read as ($\widetilde{\Delta}=0$):
 \begin{eqnarray}
\begin{split}
\mathcal{H}^E_3(t)&=\frac{1}{2}g(t)(1+\epsilon)\Big[e^{i\widetilde{\varphi}}(|00\rangle_L\langle a_1|+|11\rangle_L\langle a_2|) \\
&\quad+\textrm{H.c.}\Big],
\end{split}
 \end{eqnarray}
in which $\epsilon$ denotes the error ratio of local $\sigma_x$ error. From the Fig.  \ref{Figure6} (b), we can easily see that the gate robustness for dynamical-corrected geometric $U_2(\pi/2)$ is the best, although the maximum gate fidelity is decreased slightly, which is consistent with the case of single-logical-qubit geometric gates. In this way, by combining the dynamical-corrected universal single-logical-qubit geometric gates, we can implement the arbitrary-logical-qubit geometric quantum gates of dynamical correction based on DFS coding. It is worth noting that our two-qubit dynamically corrected geometric gates with DFS coding can also effectively resist the $\sigma_z$ error caused by the collective dephasing.

\section{Conclusion}\label{sec5}
In conclusion, we have proposed a general scheme for dynamical-corrected nonadiabatic geometric quantum gates, whose gate robustness is much superior to that of the conventional single-loop geometric scheme and even better than the two-loop composite-pulse geometric one in terms of resisting $\sigma_x$ error. Furthermore, we give the physical implementation of our scheme combined with DFS coding, and numerical results show that the implemented dynamical-corrected geometric gates can not only resist $\sigma_x$ error, but also suppress $\sigma_z$ error caused by the collective dephasing. All of these indicate that our scheme provides a promising way to achieve robust and large-scale fault-tolerant geometric quantum computation.

\begin{acknowledgements}
This work is supported by the Key-Area Research and Development Program of
GuangDong Province (grant number 2018B030326001),  the National Natural Science Foundation of China (Grant No. 12275090), Guangdong Provincial Key Laboratory (Grant No. 2020B1212060066), the Quality Engineering Project of the Education Department of Anhui Province (No.2021cyxy046), the key Scientific Research Foundation of Anhui Provincial Education Department (KJ2021A0649), Outstanding YoungTalents in College of Anhui Province (Grant No. gxyq2022059), and the high-level talent scientific research starting foundation (Grant No. 2020rcjj14).

\end{acknowledgements}

\appendix
\section{Analytic solution} \label{applxa}
To more clearly see the advantages of our dynamically corrected geometric scheme, here we give the analytic solutions of the three different geometric schemes in the main text, against local $\sigma_x$ error.

Selecting S, T and H gates as the typical examples. For the conventional single-loop geometric and the two-loop composite-pulse geometric schemes, the output fidelities $F$ of corresponding S gates suffered by $\sigma_x$ error $\epsilon$ can be calculated as, respectively,
\begin{eqnarray}
F_1^S &=&1+ \frac{1}{8}(-2+\sqrt{2})\pi^{2}\epsilon^2+\mathcal{O}(\epsilon^4), \\
F_2^S &=& 1+\frac{1}{4}\pi^2\big[-2-\sqrt{2}+\sqrt{2(2+\sqrt{2})} \nonumber \\
&\,&+\sqrt{2}\sin(\pi/8)\big]\epsilon^2+\mathcal{O}(\epsilon^4),
\end{eqnarray}
while its form is, for our dynamically corrected geometric scheme,
\begin{equation}
F_3^S=1+\frac{1}{32}(-2+\sqrt{2})\pi^{4}\epsilon^4+\mathcal{O}(\epsilon^5).
\end{equation}
Note that the forms of analytic solution for T gates are similar to that for S gates in the three different geometric schemes, and not shown here. For H gate, the target fidelities for above three geometric schemes are, respectively,
\begin{eqnarray}
F_1^H&=& 1-\frac{5}{32}\pi^2\epsilon^2+\mathcal{O}(\epsilon^4),\\
F_2^H &=& 1+\frac{1}{32}(-13+8\sqrt{2})\pi^2\epsilon^2+\mathcal{O}(\epsilon^4), \\
F_3^H &=& 1-\frac{2}{87}\pi^2\epsilon^2+\frac{1}{41}\pi^3\epsilon^3+\mathcal{O}(\epsilon^4).
\end{eqnarray}

\begin{figure}
\vspace{0.5em}
\centering
\includegraphics[width=0.95\linewidth]{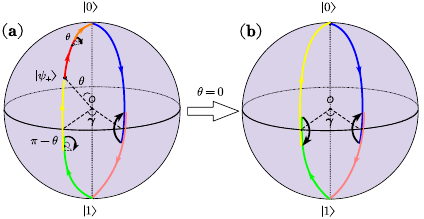}\\
\caption{The entire evolution paths under the influence of $\sigma_x$ error for (a) dynamical-corrected geometric H ($\theta=\pi/4$) and (b) S (or T) gates ($\theta=0$), respectively.}\label{Figure7}
\end{figure}

It is easy to find that the infidelities for the conventional single-loop geometric and two-loop composite-pulse geometric S (or T) gates against $\sigma_x$ error still have the same order of magnitude, i.e., are suppressed to the second order. However, the one for our dynamically corrected geometric S (or T) gate can be suppressed up to fourth order. Note that only two-loop composite-pulse geometric scheme is considered here for comparison, due to the finite coherent time. For geometric H gate, compared with the other two scheme, our scheme can further resist the $\sigma_x$ error, and is better than the one in Ref. \cite{liang2022robust}. However, it cannot be suppressed to the fourth-order. This reason is that the inserted first- and third-segment dynamical evolutions need be rotate $\theta$ and $\pi-\theta$ angles around the axes $\textbf{n}_1=(\theta/2,\phi)$ and $\textbf{n}_3=(\frac{\pi+\theta}{2},\phi)$ at $\theta\neq0$, respectively, whose corresponding $\sigma_x$ errors can only be partially eliminated. It is different from the inserted second-segment dynamical evolution that is rotated $\pi$ angle around the axis of $\textbf{n}_2=(\pi/2, \phi+\gamma)$, which can nearly completely eliminate the $\sigma_x$ error. Thus, the overall effect is to partially correct the $\sigma_x$ error in the case of $\theta\neq0$. As shown in Fig. \ref{Figure7}, when $\theta=0$, the asymmetric dynamical correction in (a) would become the symmetric one in (b). The latter can almost completely eliminate the $\sigma_x$ error.
\section{Robustness comparison for $\theta\neq0$} \label{applxb}
\begin{figure}
 \vspace{0.5em}
\centering
\includegraphics[width=0.95\linewidth]{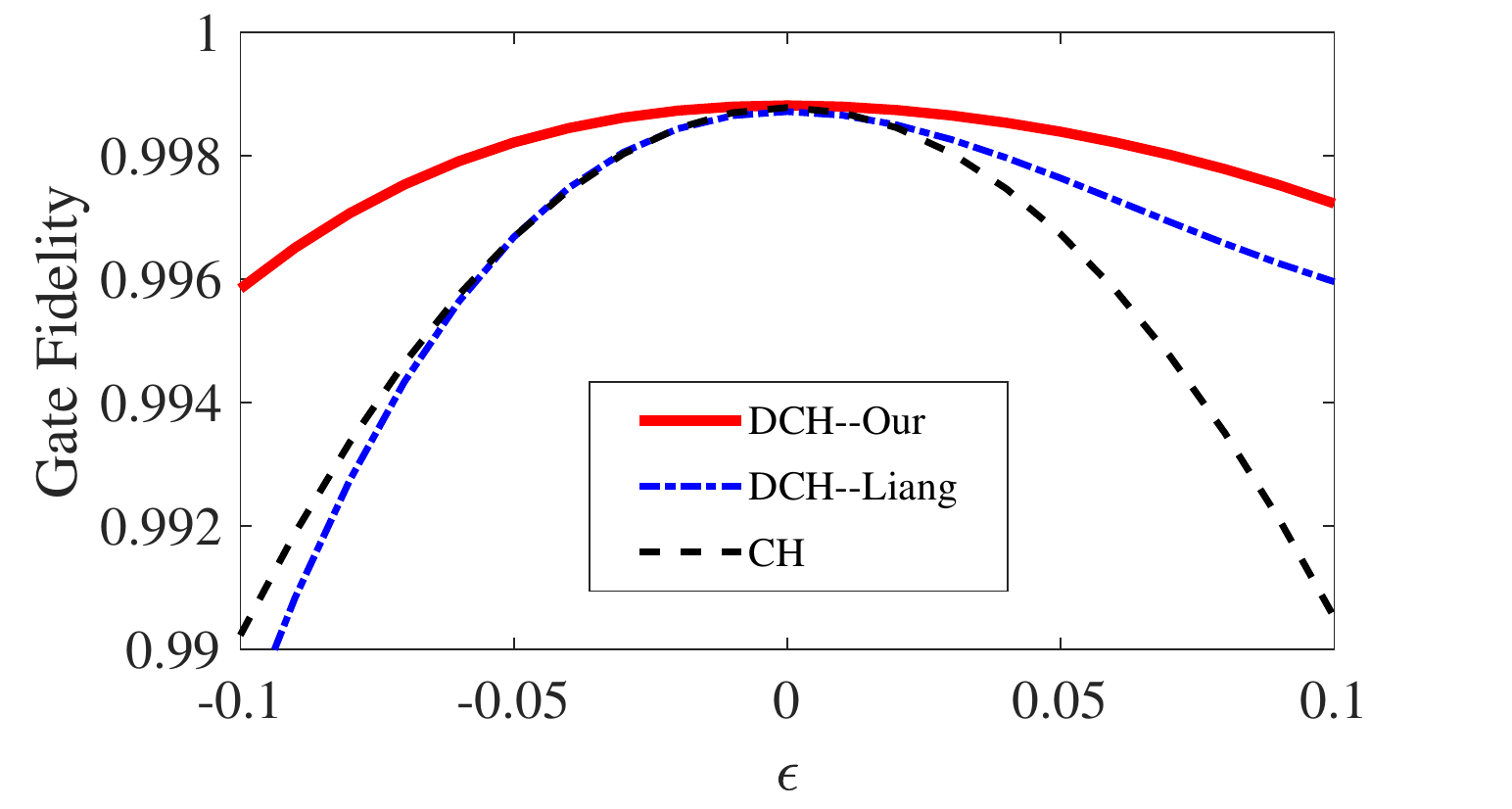}\\
\caption{Robustness comparison of geometric Hadamard gate in resisting $\sigma_x$ error with decoherence, among the two-loop composite-pulse geometric scheme (black-dashed line), the optimization geometric one in Ref. \cite{liang2022robust} (blue-dashed-dotted line) and our scheme (red-solid line).}\label{Figure8}
\end{figure}
\begin{figure}
 \vspace{0.5em}
\centering
\includegraphics[width=0.95\linewidth]{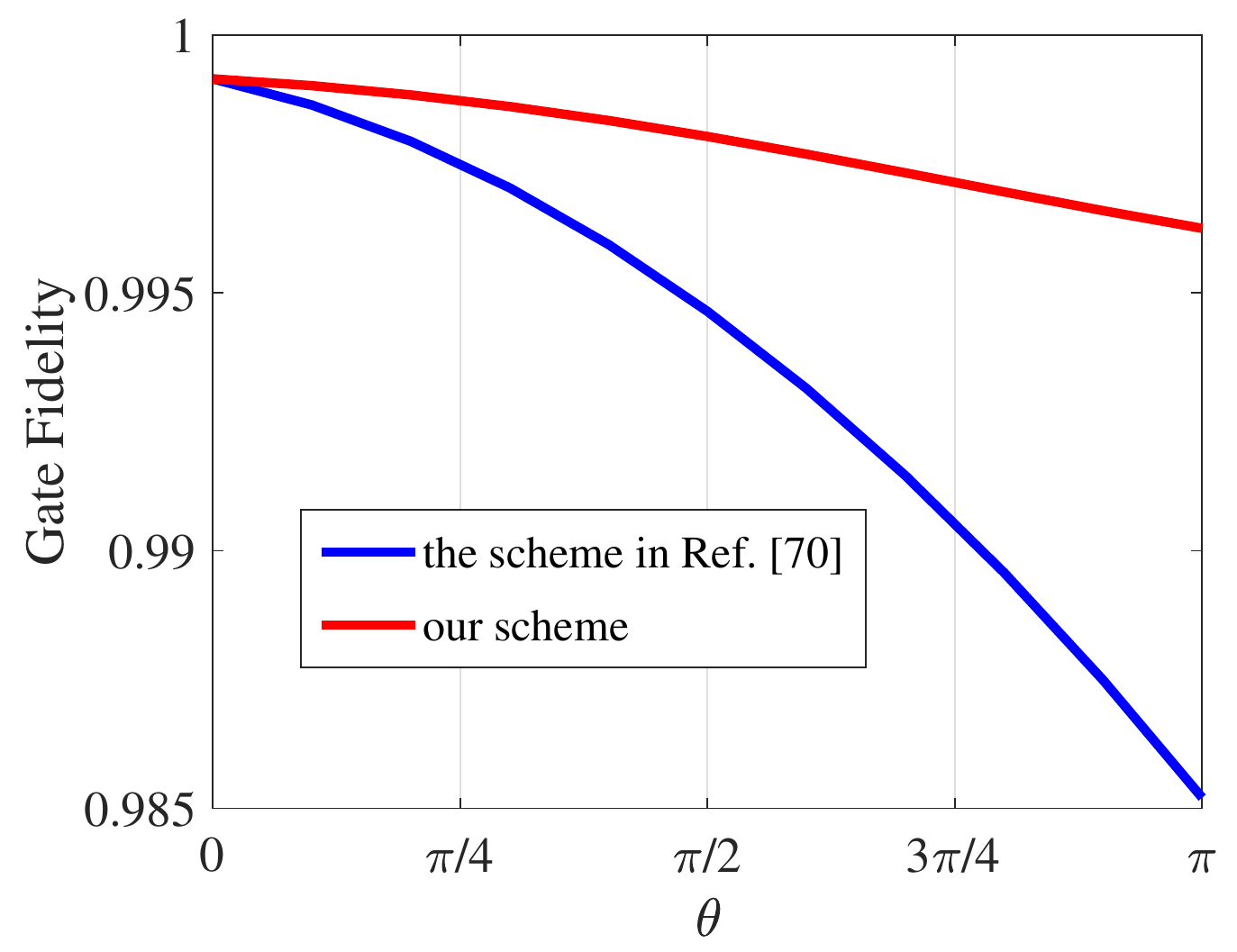}\\
\caption{The gate fidelities versus parameter $\theta$ for our scheme (red line) and that in Ref. \cite{liang2022robust} (blue line), respectively, where $\epsilon=-0.1$, $\phi=0$ and $\gamma=-\pi/2$. }\label{Figure9}
\end{figure}

To demonstrate gate robustness advantage of our dynamical-corrected geometric scheme under $\theta\neq0$, taking the geometric Hadamard gate $(\theta=\pi/4)$ as the typical example, we numerically compare the gate robustness between the scheme in Ref. \cite{liang2022robust} and our scheme under decoherence, which is shown in Fig. \ref{Figure8}. Moreover, the gate robustness against $\sigma_x$ error for the two-loop composite-pulse geometric Hadamard gate is also plotted in Fig. \ref{Figure8} for comparison. On the one hand, it can be found that the gate robustness for our scheme is stronger than that in Ref. \cite{liang2022robust} and the two-loop composite-pulse geometric scheme, especially for $\epsilon<0$. On the other hand, compared with the one for two-loop composite-pulse geometric scheme, the gate robustness of the optimization scheme in Ref. \cite{liang2022robust} is less weaker under $\epsilon<0$, although it is more robust under $\epsilon>0$. The reason of the considerable asymmetry is mainly that it can only dynamically optimize two of the more general three-segment pulses in the conventional single-loop geometric scheme, apart from a longer time. On the contrary, our dynamical-corrected geometric scheme can focally optimize them, respectively, and be regarded as a more general method. In addition, we further demonstrate that the robustness advantage of our scheme is increasing with the value of $\theta$ increased. As shown in Fig. \ref{Figure9}, we plot the gate fidelity versus parameter $\theta$ for our scheme and that in Ref. \cite{liang2022robust}, respectively, where the $\epsilon$ is set as $-0.1$, the others parameters are same as previous sections.

\bibliographystyle{apsrev4-1}
\bibliography{ref}

\end{document}